\documentclass[12pt]{article}
\usepackage{amssymb}
\newtheorem{definition}{Def}
\newtheorem{theorem}{Th}

\begin{document}

\begin{titlepage}

\begin{center}

\vskip 15cm

{\large\bf ON THE COMPATIBILITY BETWEEN THE MARKOV PROPERTY AND THE QUANTUM JUMP}
\vskip 1cm
{\bf D. Salgado}
\vskip 0.4cm
 Departamento de F\'{\i}sica Te\'{o}rica, Universidad Aut\'{o}noma de Madrid \\
Ciudad Universitaria de Cantoblanco, 28049 Cantoblanco (Madrid), Spain

\vskip 0.7cm

{\bf J.L. S\'{a}nchez-G\'{o}mez}
\vskip 0.4cm
Departamento de F\'{\i}sica Te\'{o}rica, Universidad Aut\'{o}noma de Madrid \\
Ciudad Universitaria de Cantoblanco, 28049 Cantoblanco (Madrid), Spain
\end{center}

\date{\today}

\vskip 2cm

\begin{center}
{\bf Abstract}
\end{center}
\begin{quote}

 A brief analysis of the compatibility between the quantum jump and the Markov property for quantum systems described by a stochastic evolution scheme is presented.

\end{quote}

\end{titlepage}


\section{Introduction}
\label{Intro}

The result presented here is the incompatibility, within the stochastic evolution scheme assumption, of the Markovian feature of a microscopic system and the quantum jump to which such systems are subjected (e.g. in a measurement process). These evolution schemes do not only show interest from a fundamental standpoint in quantum mechanics (cf., e.g., \cite{CSL}), but also in quantum optics (cf. \cite{plenio} and references therein) where modern technology begins to offer the possibility of monitoring single systems. This reinforces the interest on the stochastic methods applied to Hilbert spaces. In the following lines, the concept of stochastic evolution scheme is briefly set forth, the notion of quantum jump is succintly discussed and the announced incompatibility theorem is proved. A short discussion upon the entailed physical situation and its more direct consequences is also included.


\section{Quantum Stochastic Evolution Schemes}
\label{QSES}
We shall understand as Stochastic Evolution Schemes (SES) those evolution models which are non-deterministic in the sense that, given the state of a system at any instant t, we cannot predict with absolute certainty its state at a time later than t, but at most the probability of evolution towards one or other state. It must be noticed that the mathematical form in which the state of physical system should be described has not still been specified. We shall center ourselves on quantum systems, hence the latter will be determined by vectors belonging to a Hilbert space. The conjuction of both a stochastic and a hilbertian structure is not a mathematically difficult task, provided the global definition of stochastic process is kept in mind. Thus, we define a Quantum Stochastic Evolution Scheme (QSES) as a measurable application from a probability space $\left(\Omega,\mathfrak{A},P\right)$ on the set of mappings from $\mathbb{R}^{+}$ (standing for time) on the Hilbert space $\mathcal{H}$ (the state space of the system). Formally\\

\begin{displaymath}
\begin{array}{cccc}
\Psi:&\left(\Omega,\mathfrak{A},P\right)&\longmapsto& \mathcal{H}^{\mathbb{R}^{+}}\\
 &\omega&\longmapsto&\Psi(\omega)=\psi_{t}(\omega)\\
\end{array}
\end{displaymath}

where $\psi_{t}(\cdot)$ denotes a mapping from $\mathbb{R}^{+}$ onto $\mathcal{H}$ (usually denoted $\Psi(t)$ in orthodox quantum mechanics).\bigskip   

As a consequence of the metric structure of $\mathcal{H}$, it is always possible to define a $\sigma$-algebra with respect to which $\Psi$ is measurable. The usual concepts appearing in the ordinary theory of stochastic processes are still valid. In particular, we may carry on talking of the transition probabilities. Thus, we establish the following definition\bigskip 

\begin{definition}
Let $s\leq t\in \mathbb{R}^{+}$ and $\psi,\phi\in\mathcal{H}$. We call \emph{transition probability} asociated to the QSES to the quantity $P(s,\phi,t,\psi)$ defined by 
\begin{displaymath}
P(s,\phi,t,\psi)\equiv P\left(\Psi_{t}=\psi|\Psi_{s}=\phi\right)
\end{displaymath}
\end{definition}

This concept is analogous to the usual probability of Markov chains in continuous time. The time homogeneity condition (physically evident, on the other hand) is assumed from the beginning:

\begin{definition}
A QSES is \emph{homogeneous} if its transition probability is stationary, i.e., 
\begin{displaymath}
P(s+u,\phi,t+u,\psi)=P(s,\phi,t,\psi)
\end{displaymath}
for all $u$ such that $0\leq s+u\leq t+u$.\\
\end{definition}

This property enables us to speak of a transition probability in a time $t$, given by
\begin{displaymath}
P(t;\phi,\psi)=P(0,\phi,t,\psi)=P(u,\phi,t+u,\psi)
\end{displaymath}
The quantity $P(t;\phi,\psi)$ will be our basic tool to obtain the desired result. In an obvious way, it satisfies the following relations:

\begin{eqnarray} \label{pos}
P(t;\phi,\psi)\geq 0 & \forall \psi,\phi \in \mathcal{H},\forall t \in [0,\infty) \\ \label{norm}
\int_{\mathcal{H}}P(t;\phi,\psi)\mu(d\psi)=1 & \forall \phi \in \mathcal{H}
\end{eqnarray}

where $\mu$ denotes the Lebesgue measure on $\mathcal{H}$. The conditions (\ref{pos}) and (\ref{norm}) briefly state that $P(t;\phi,\psi)$ is for each $t\in\mathbb{R}^{+}$ a stochastic matrix. The Markov property is equally stated in this formalism: 

\begin{definition}
A QSES is said to be Markovian if it satisfies
\begin{displaymath}
\int_{\mathcal{H}}P(t;\phi,\psi)P(s;\psi,\varphi)\mu(d\psi)=P(t+s;\phi,\varphi)\quad \forall\phi,\varphi\in\mathcal{H},\forall t,s\in\mathbb{R}^{+}
\end{displaymath}
\end{definition}

This assumption will be the central objective of our analysis. The fundamental result we need is the following
\begin{theorem}
Let $P(\cdot;\phi,\psi)$ be a stochastic transition matrix corresponding to a Markovian QSES. Then $P(\cdot;\phi,\psi)$ is continuous in $(0,\infty)$ for all $\phi,\psi\in\mathcal{H}$ if and only if the following limit exists
\begin{displaymath}
\lim_{t\to0^{+}}P(t;\phi,\psi)=g(\phi,\psi)
\end{displaymath}
\end{theorem}
where $g(\phi,\psi)$ satisfies 
\begin{displaymath}
\begin{array}{rclc}
g(\phi,\psi) & \geq & 0 & \forall\phi,\psi\in\mathcal{H}\\
\int_{\mathcal{H}}g(\phi,\psi)\mu(d\psi) & \leq & 1 & \forall\phi\in\mathcal{H}\\
g(\phi,\psi) & = & \int_{\mathcal{H}}g(\phi,\varphi)g(\varphi,\psi)\mu(d\varphi) & \forall\phi,\psi\in\mathcal{H}
\end{array}
\end{displaymath}

This theorem is but a translation of the corresponding known theorem for Markov chains in continuous time (cf. \cite{Chung}). We shall outline its proof and relegate minor details to the appendix. Notice should be taken of the generality enabled by the function $g(\phi,\psi)$ whose expression has not been detailed. The utility of this result rests on the possibility of checking the continuity of a stochastic matrix at every point through the study of a simple limit at the origin. The only assumed hypothesis is the Markov property.

Proof.\\ 
$(\Rightarrow)$ Let us suppose that $P(t;\phi,\psi)$ is continuous in $(0,\infty)$. By Bolzano-Weierstrass theorem it is possible to find sequences $\{t_n\}$ and $\{t'_{n}\}$ such that 
\begin{eqnarray}
P(t;\phi,\psi) & = &\lim_{n\to\infty}P(t+t_{n};\phi,\psi)\nonumber\\
P(t';\phi,\psi) & = &\lim_{n\to\infty}P(t'+t'_{n};\phi,\psi)\nonumber
\end{eqnarray}
Let us call 
\begin{eqnarray}
u(\phi,\psi) & \equiv &\lim_{n\to\infty}P(t_{n};\phi,\psi)\nonumber\\
u'(\phi,\psi) & \equiv &\lim_{n\to\infty}P(t'_{n};\phi,\psi)\nonumber
\end{eqnarray}
Our goal is to establish that $u(\phi,\psi)=u'(\phi,\psi)$ for all $\phi,\psi\in\mathcal{H}$. By making use of the continuity hypothesis, the Markov property and some measure-theoretic usual theorems, it can be easily shown that the following two relations are simultaneously fulfilled (cf. Appendix):  

\begin{eqnarray} \label{mayor}
u(\phi,\psi) & \geq & \int_{\mathcal{H}}u'(\phi,\varphi)u(\varphi,\psi)\mu(d\varphi) \\ \label{igual}
u'(\phi,\psi) & = & \int_{\mathcal{H}}u'(\phi,\varphi)u(\varphi,\psi)\mu(d\varphi) 
\end{eqnarray}

whence by the symmetry and arbitrariness of  $\phi$ and $\psi$, the equality $u(\phi,\psi)=u'(\phi,\psi)$ for all $\phi,\psi\in\mathcal{H}$ is derived. 

$(\Leftarrow)$ Let us now suppose that there exists a unique limit $u(\phi,\psi)$ when $t\to0$. Then

\begin{eqnarray}
\lim_{n\to\infty}P(t+t_{n};\phi,\psi) & = & \lim_{n\to\infty}\int_{\mathcal{H}}P(t;\phi,\varphi)P(t_{n};\varphi,\psi)\mu(d\varphi)= \nonumber \\
& = & \int_{\mathcal{H}}P(t;\phi,\varphi)u(\varphi,\psi)\mu(d\varphi)=P(t;\phi,\psi) \nonumber
\end{eqnarray}
which implies that $P(t;\phi,\psi)$ is right continuous. But it is elementary to show that a right continuous function has at most a denumerable set of discontinuities, hence it is measurable. Then we are left with the task of proving that a Markovian stochastic transition matrix which has a denumerable set of discontinuities is continuous, result which is established in the appendix.


\section{The quantum jump}
\label{Qjump}
Undoubtedly the quantum jump is one of the most controversial aspects of orthodox quantum mechanics. We are, for the analysis we set forth here, interested in the following aspect of this phenomenon. According to Mittelstaedt's schematic representation (cf. \cite{Mitt}), the measurement process can be divided into three stages, namely, preparation, premeasurement and objectification and reading. In particular we are interested in the fact that during premeasurement $0\leq t\leq t'$, the composite system system+apparatus evolves unitarily following quantum-mechanical laws. At the very instant $t'$, the objectification and reading stage begins. It is in this transition where the core of the measurement problem is rooted and where the origin of the reduction postulate is located. We will partially assume that such a reduction takes place, i.e., at $t'$ the state vector transforms instantaneously (\emph{jumps}) into another state vector (autostate of the measured observable). However, we do not enter into considerations about the origin or the factors of that jump, not even we attempt to interpret it. We only assume as a hypothesis that there exist physical situations in which the state vector instantaneously jumps to another vector. It is even permitted that the difference between such vectors be of non-null norm. We must now translate these ideas into the language of QSES's. We then say that a QSES reflects the quantum jump if its transition matrix satisfies    

\begin{displaymath}
P(t;\phi,\psi)=\left\{\begin{array}{l}
                     \left\{\begin{array}{ll}
                           1 & \textrm{  if  } 0\leq t\leq t' \textrm{  and  } \psi=U(t)\phi\\
                           0 & \textrm{  si  } 0\leq t\leq t' \textrm{  and  } \psi\neq U(t)\phi
                           \end{array}\right.\\
                       \\
                      h(\phi,\psi)(\neq 0) \textrm{  if  } t>t' \textrm{  and it is possible  }\\
                       ||\psi-\phi||>\epsilon \textrm{  for some  } \epsilon>0
                      \end{array}\right.
\end{displaymath}

where U(t) is the usual quantum-mechanical evolution operator.\bigskip

Notice that $h(\psi,\phi)=|(\phi,\psi)|^{2}$ should be expected in order to reproduce the reduction postulate. We only claim that after an unitary evolution during the premeasurement and being arrived at the instant of objectification and reading, the system jumps with finite non-null probability. This hypothesis is clearly of physical nature and its relationship with the Markovianity of the QSES constitutes the central aim of this note. The more relevant mathematical property involved is the discontinuity of $P(t;\phi,\psi)$, which may be showed using usual Calculus techniques applied to the definition of a quantum jump in a QSES. 
 

\section{The quantum jump and the Markov condition}
\label{ContMarkov}
To confront the two previous hypothesis (Markovian QSES and quantum jump) is not an exceedingly complicated task provided we make use of the preceding results. We in no case adopt a priori attitudes, but only study the compatibility of both ideas, which we present in the form of the following

\begin{theorem}
Let S be a quantum system described by a time-homogeneous QSES. If S is subjected to quantum jumps, then its QSES is non-Markovian.\\
\end{theorem}

Proof. The proof, though elementary, requires the use of a physical hypothesis which lately we shall comment in greater detail. Let us suppose that the mentioned QSES is Markovian, then its stochastic matrix is continuous in $\mathbb{R}^{+}$ if and only if the limit $\lim_{t\to0}P(t;\phi,\psi)$ exists for all $\phi,\psi \in\mathcal{H}$. This limit, due to physical assumptions, exists and indeed amounts to  

\begin{equation} \label{estand}
\lim_{t\to0}P(t;\phi,\psi)=\left\{\begin{array}{cc}
0 & \textrm{si }\phi\neq\psi\\
1 & \textrm{si }\phi=\psi
\end{array}\right.
\end{equation}

 Then $P(t;\phi,\psi)$ is continuous for all $t$ and in particular at the instant $t'$ in which the jump takes place, contrary to the assumed existence of such a jump.

The core of the proof is undoubtedly the question about the existence of the limit at the origin. The validity of such an assumption, though apparently trivial, we believe, deserves careful discussion.


\section{Discussion}
\label{Discu}
Firstly we will consider the question about the existence of the limit at the origin, distinguishing between physical and mathematical aspects. We shall refer to the existence of such a limit with the value formerly assigned by eq. (\ref{estand}) as \emph{standard condition} o \emph{standarization}, in clear analogy with Markov chains. In orthodox quantum mechanics, in which the state vector evolves in a deterministc fashion, this question is positively solved by means, e.g., of the imposition of the usual initial condition on the evolution operator $U(t_{o},t_{o})=I$. The physical interpretation is immediate: the state vector of a quantum physical system does not change if time has hardly elapsed. Furthermore, standarization appears as a hypothesis \emph{assumed on physical grounds} in the study of the quantum Zeno paradox (cf. \cite{Zeno}).\bigskip 

However, we should now discuss if standarization is kept when the mathematical character of the state is changed, i.e., instead of being represented by a vector in a Hilbert space, we let it be represented by a Hilbert-space-valued stochastic process. We think that there exist notably suggestive reasons to claim that the standard condition is satisfied. Let us consider, e.g., the open quantum system formalism. There the system evolution is given by an operator semigroup which satisfies, among other properties, standarization (cf. \cite{Davies}). And this is so even despite the enviromental uncontrollable influence.\bigskip

In connection with the previously established theorem, we should remember that the standard condition is not necessary, since, according to section \ref{QSES}, it is only required the existence of the limit at the origin independently of its value.\bigskip

We believe equally important to draw our attention on the implications of the theorem of the foregone section. It does not establish the impossibility of the quantum jump, nor does it deny the Markovian character of the stochastic evolution of a quantum systm. We understand the result is: \textbf{Assumed} the description of a quantum system by means of a homogeneous QSES, \textbf{if} the system exhibits quantum jumps, \textbf{then} the QSES cannot be Markovian. What attitudes can be adopted before such a situation? Firstly, we can neglect the possibility of mathematically representing a quantum jump through an $\mathcal{H}$-valued stochastic process. This solution is doubtlessly the sharpest, but in our opinion too restrictive. Secondly, it can be claimed that the quantum jump is not real, i.e., it does not take place and that the evolution of a quantum system, though stochastic, is continuous. This is the hypothesis adopted, e.g., in the CSL theory (cf. \cite{CSL}). Nonetheless, it is also possible that the Markov condition not be satisfied even maintaining continuity in the evolution, as in, e.g., \cite{NonMarkQSD}. Finally, the option is left of admitting every hypothesis in the theorem with the subsequent consequences. This alternative has not been studied profoundly yet.

\section*{Acknowledgements}
One of us (D.S.) must acknowledge the support of the CAM Education Council under grant BOCAM-20/8/99.

\section{Appendix}
To establish (\ref{mayor}) we must apply the continuity hypothesis, the Markov property, Fatou lemma and the definition of $u(\phi,\psi)$ in the following way:

\begin{eqnarray}
P(t;\phi,\psi) & = & \liminf_{n\to\infty}P(t'_{n}+t;\phi,\psi)= \nonumber \\
& = & \liminf_{n\to\infty}\int_{\mathcal{H}}P(t'_{n};\phi,\varphi)P(t;\varphi,\psi)\mu(d\varphi)\geq \nonumber \\
& \geq & \int_{\mathcal{H}}u'(\phi,\varphi)P(t;\varphi,\psi)\mu(d\varphi) \nonumber
\end{eqnarray}
Since this is fulfilled for all $t\in(0,\infty)$, it will be in particular satisfied for each $t_{n}$, which enables us to write, making use again of Fatou lemma:
\begin{displaymath}
\liminf_{n\to\infty}P(t_{n};\phi,\psi)=u(\phi,\psi)\geq \int_{\mathcal{H}}u'(\phi,\varphi)u(\varphi,\psi)\mu(d\varphi)
\end{displaymath}
which is the sought relation.\bigskip

To settle eq. (\ref{igual}) we must show a few partial results:

\begin{description}
\item[1º)] Applying the continuity hypothesis, the Markov property and the dominated convergence theorem, we obtain for all $t\in\mathbb{R}^{+}$:
\begin{eqnarray}
P(t;\phi,\psi) & = & \lim_{n\to\infty}P(t+t_{n};\phi,\psi)= \nonumber\\
 & = & \lim_{n\to\infty}\int_{\mathcal{H}}P(t;\phi,\varphi)P(t_{n};\varphi,\psi)\mu(d\varphi) = \nonumber\\
 & = & \int_{\mathcal{H}}P(t;\phi,\varphi)u(\varphi,\psi)\mu(d\varphi) \nonumber
\end{eqnarray}\\
\item[2º)] Making use of property (\ref{norm}) of stochastic matrices, of the just obtained result and of Fubini theorem, we may write:
\begin{eqnarray}
1 & = & \int_{\mathcal{H}}P(t;\phi,\psi)\mu(d\psi)= \nonumber\\
  & = & \int_{\mathcal{H}}\left[\int_{\mathcal{H}}P(t;\phi,\varphi)u(\varphi,\psi)\mu(d\varphi)\right]\mu(d\psi)= \nonumber\\
  & = & \int_{\mathcal{H}}P(t;\phi,\varphi)\left[\int_{\mathcal{H}}u(\varphi,\psi)\mu(d\psi)\right]\mu(d\varphi) \nonumber
\end{eqnarray}
whence it is deduced that if $\int_{\mathcal{H}}u(\varphi,\psi)\mu(d\psi)<1$, then $P(t;\phi,\varphi)=0$ a.e. for all $\phi\in\mathcal{H}$ and for all $t\in\mathbb{R}^{+}$. In particular, it is satisfied for all $t'_{n}$, then 
\begin{displaymath}
\textrm{If  }\int_{\mathcal{H}}u(\varphi,\psi)\mu(d\psi)<1,\textrm{then  } u'(\phi,\varphi)=0 \textrm{ a.e. } \forall\phi\in\mathcal{H}.
\end{displaymath}\\
\item[3º)] Again using the same techniques as before, we may write:
\begin{eqnarray}
u'(\phi,\psi) & = & \liminf_{n\to\infty}P(t'_{n};\phi,\psi)= \nonumber \\
 & = & \liminf_{n\to\infty}\int_{\mathcal{H}}P(t'_{n};\phi,\varphi)u(\varphi,\psi)\mu(d\varphi)\geq \nonumber\\
 & \geq & \int_{\mathcal{H}}u'(\phi,\varphi)u(\varphi,\psi)\mu(d\varphi) \nonumber
\end{eqnarray}
\end{description}
With these and again the already used results, we arrive at
\begin{eqnarray}
\int_{\mathcal{H}}u'(\phi,\psi)\mu(d\psi) & \geq &\int_{\mathcal{H}}\left[\int_{\mathcal{H}}u'(\phi.\varphi)u(\varphi,\psi)\mu(d\varphi)\right]\mu(d\psi)= \nonumber \\
 & = & \int_{\mathcal{H}}u'(\phi,\varphi)\left[\int_{\mathcal{H}}u(\varphi,\psi)\mu(d\psi)\right]\mu(d\varphi)= \nonumber \\
 & = & \int_{\mathcal{H}}u'(\phi, \varphi) \mu(d\varphi) \nonumber
\end{eqnarray}
whence straightforwardly it is obtained the second sought relation.\bigskip

For the sufficiency, we present the deduction of continuity of $P(t;\phi,\psi)$ from its measurability in the form of a 

\begin{theorem}
Let $P(t;\phi,\psi)$ be measurable for all $\phi,\psi\in\mathcal{H}$, then $P(t;\phi,\psi)$ is continuous in $(0,\infty)$.
\end{theorem}

The proof will be established by parts. Firstly we shall show that the expression
\begin{equation} \label{serie}
\int_{\mathcal{H}}\left|P(t+h;\phi,\psi)-P(t;\phi,\psi)\right|\mu(d\psi)
\end{equation}
is a non-increasing function of $t$. Secondly, we will show that (\ref{serie}) converges uniformly to $0$ when $h\to0$, where $t\geq\delta>0$. Thus it is elementarily deduced that $P(t;\phi,\psi)$ is uniformly continuous in $[\delta,\infty)$ for all $\delta>0$ or equivalently in $(0,\infty)$. 
1) Using Markov property and Fubini theorem it is showed that for $0<s<t$
\begin{displaymath}
\int_{\mathcal{H}}\left|P(t+h;\phi,\psi)-P(t;\phi,\psi)\right|\mu(d\psi) = \nonumber
\end{displaymath}
\begin{displaymath}
 = \int_{\mathcal{H}}\left|\int_{\mathcal{H}}\left[P(s+h;\phi,\varphi)-P(s;\phi,\varphi)\right]P(t-s;\varphi,\psi)\mu(d\varphi)\right|\mu(d\psi)\leq  \nonumber
\end{displaymath}
\begin{displaymath}
\leq\int_{\mathcal{H}}\left|P(s+h;\phi,\varphi)-P(s;\phi,\varphi)\right|\mu(d\varphi)\int_{\mathcal{H}}P(t-s;\varphi,\psi)\mu(d\psi)=
\end{displaymath}
\begin{displaymath}
\int_{\mathcal{H}}\left|P(s+h;\phi,\varphi)-P(s;\phi,\varphi)\right|\mu(d\varphi)
\end{displaymath}
And the first part of the proof is established.
2) If now the measurability hypothesis of $P(t;\phi,\psi)$ is introduced, we can integrate between $0$ and $\delta\leq t$ to obtain the bound
\begin{displaymath}
 \int_{\mathcal{H}}\left|P(t+h;\phi,\psi)-P(t;\phi,\psi)\right|\mu(d\psi)\leq
\end{displaymath}
\begin{displaymath}
\leq\int_{\mathcal{H}}\frac{1}{\delta}\left[\int_{0}^{\delta}\left|P(s+h;\phi,\varphi)-P(s;\phi,\varphi)\right|ds\right]\mu(d\varphi)
\end{displaymath}
where if $0\leq h\leq \delta$ the second term is dominated by 
\begin{displaymath}
\int_{\mathcal{H}}\frac{2}{\delta}\left[\int_{0}^{2\delta}P(s;\phi,\varphi)ds\right]\mu(d\varphi)
\end{displaymath}
hence uniform convergence is established. Now, it is elementary to show that for each $\phi,\psi\in\mathcal{H}$
\begin{eqnarray}
\lim_{h\to0}\int_{0}^{\delta}\left|P(s+h;\phi,\psi)-P(s;\phi,\psi)\right|ds & = & 0 \nonumber
\end{eqnarray}
which together with the uniform convergence implies the second part of the proof.

The properties of $g(\phi,\psi)$ referred in the text are easily deduced from the several relations obtained throughout the proof. 


\end{document}